\documentstyle[aps,prl,epsfig] {revtex}

\begin{document}
\draft


\title { The rate of entropy increase at the edge of chaos}

\author{   Vito Latora $^{(a)}$   }

\address{Center for Theoretical Physics, Laboratory for Nuclear Sciences 
and Department of Physics, 
\\ Massachusetts Institute of Technology, Cambridge, Massachusetts 02139, USA\\
Department of Physics, Harvard University, Cambridge, Massachusetts, 02138, 
USA\\
and\\
Dipartimento di Fisica,   Universit\'a di Catania, \\
and Istituto Nazionale di Fisica Nucleare, Sezione di Catania 
Corso Italia 57, I-95129 Catania, Italy\\} 

\author{Michel Baranger$^{(b)}$}

\address{Center for Theoretical Physics, Laboratory for Nuclear Sciences and
Department of Physics,\\
Massachusetts Institute of Technology, 
Cambridge, Massachusetts 02139, USA\\}

\author{  Andrea Rapisarda $^{(c)}$ }

\address{Dipartimento di Fisica,   Universit\'a di Catania, \\
and Istituto Nazionale di Fisica Nucleare, Sezione di Catania 
Corso Italia 57, I-95129 Catania, Italy }

\author{  Constantino Tsallis  $^{(d)}$  }

\address{Centro Brasileiro de Pesquisas Fisicas, 
Xavier Sigaud 150, 22290-180, 
Rio de Janeiro-RJ, Brazil,\\
Physics Department, University of North Texas, P.O.     
     Box 311427, Denton, Texas 76203, USA\\
and\\
Department of Mechanical Engineering, Massachusetts Institute of Technology, Rm. 3-164,
Cambridge, Massachusetts 02139, USA}

\date{\today}
\maketitle


\begin{abstract}

Under certain conditions, the rate of increase of the statistical entropy of a
simple, fully chaotic, conservative system is known to be given by a single
number, characteristic of this system, the Kolmogorov--Sinai entropy rate.
This connection is here generalized to a simple dissipative system, the
logistic map, and especially to the chaos threshold of the latter, the edge of
chaos.  It is found that, in the edge--of--chaos case, the usual
Boltzmann--Gibbs--Shannon entropy is not appropriate.  Instead, the
non--extensive entropy  $
S_q\equiv \frac{1-\sum_{i=1}^W p_i^q}{q-1}\;
$, must be used. The latter  contains  a parameter $q$, the entropic index 
which must be given 
a special value $q^*\ne 1$ (for  $q= 1$ one recovers the usual entropy)
characteristic of the edge--of--chaos under consideration. 
The {\it same} $q^*$ enters also in the description of the sensitivity to initial
conditions, as well as in that of the multifractal spectrum of the attractor.

\end{abstract}


 {\em Final Version accepted for publication in Physics Letters A}


\bigskip

The connection between chaos and thermodynamics has been receiving increased
attention lately.  A review of the central ideas can be found in \cite{review}.
Recent studies have focused on both {\it conservative} (classical long--range
interacting many--body Hamiltonians\cite{anteneodo}, low--dimensional
conservative maps\cite{vm}) and {\it dissipative} (low--dimensional
maps\cite{tsallislyra,tsallisplastino}, many--body self--organized
criticality\cite{SOC}, symbolic sequences\cite{buiatti}) systems.  In ref.
\cite{vm} the connection between the Kolmogorov--Sinai (KS) entropy rate and the
statistical entropy (or thermodynamic entropy) was brought out for simple
conservative systems.  In ref. \cite{tsallislyra} the non--extensive entropy
introduced some years ago by one of us \cite{tsallis1} was shown to be the
relevant quantity at the chaos threshold ({\it the edge of chaos}).  This
entropy contains a parameter $q$ which has been called {\it the entropic index}
and it reduces to the usual Boltzmann--Gibbs--Shannon (BGS) 
entropy when $q = 1$. In refs. \cite{tsallislyra,tsallisplastino} 
$q$ was determined in two completely different ways:
from the sensitivity to initial conditions and from the multifractal spectrum (using $1/(1-q) = 1/\alpha_{min} - 1/\alpha_{max}$, where $\alpha_{min}$ and $\alpha_{max}$ are respectively  the lower and upper values where the multifractal function $f(\alpha)$ vanishes).

The purpose of this letter is to extend the work of \cite{vm} to a dissipative
case and to focus especially on the edge of chaos.  The results are:
1) In the chaotic regime the linear rate of the BGS entropy gives the KS 
entropy;
2) At the edge of chaos, the non--extensive entropy for one  particular 
value $q\ne 1$  grows   linearly with time, as did the usual entropy in \cite{vm};
3) The value of $q$ thus determined at the edge of chaos is identical with that found with the two
   other independent methods respectively used in refs. 
\cite{tsallislyra} and \cite{tsallisplastino}.

The dissipative system chosen is the simplest possible: the logistic map, a
nonlinear one--dimensional dynamical system described by the iterative
rule \cite{review}:
\begin{equation}
x_{t+1}=1-ax_t^2\;\;\;(-1 \le x_t \le1; \;0\le a \le2;\;t=0,1,2,...)\; .
\end{equation}
It has chaotic behavior (with a positive Lyapunov exponent) for most of the
values of the control parameter $a$ above the critical value $a_c\equiv
1.40115519...$.  This critical value marks the edge of chaos.

For convenience, we recall here the definition of the non--extensive entropy
\cite{tsallis1}.  If the phase space $\cal{R}$ has been divided into $W$ cells
of equal measure, and if the probability of being in cell $i$ is $p_i$, we
define the entropy $S_q$ by
\begin{equation}
S_q\equiv \frac{1-\sum_{i=1}^W p_i^q}{q-1}\;\;\;\;(q \in \cal{R})\; .
\label{entropy}
\end{equation} 
For $q = 1$ this is $S_1=-\sum_{i=1}^Wp_i\;\ln\;p_i$, the usual entropy.
This generalized entropy was proposed a decade ago \cite{tsallis1} to allow
statistical mechanics to cover certain anomalies due to a possible
(multi)fractal structure of the relevant phase space (for example, whenever we
have long--range interactions, long--range microscopic memory, multifractal
boundary conditions, some dissipative process, etc). A review of the existing
theoretical, experimental and computational evidence and connections is now
available\cite{tsallis2} (very recent verifications in fully developed turbulence and in electron-positron annihilation producing hadronic jets are exhibited in \cite{beck1} and \cite{beck2} respectively).
We also recall the two main results 
of refs. \cite{tsallislyra,tsallisplastino}.
The first one \cite{tsallislyra} concerns the 
sensitivity to initial conditions.  In a truly
chaotic system, the separation between nearby trajectories, suitably averaged
over phase space, $\xi (t)$ diverges in time like 
the exponential $\exp{(\lambda_1 t)}$,
where $\lambda_1>0$ is the Lyapunov exponent.  At the edge of chaos, on the
other hand, the upper bound  growth of the separation follows 
a power law which may be written
\begin{equation}
   \xi(t) \;\propto \; [1+(1-q)\lambda_q t]^{\frac{1}{1-q}}\;\;\;(q \in \cal{R})
\label{monster}
\end{equation}
in terms of a certain parameter $q$.  The exponential is recovered in the limit
$q=1$. For the logistic map at the edge of chaos, the Lyapunov exponent
$\lambda_1$ is found to vanish, but the growth of the separation is fitted well
\cite{nota1} by eq. (\ref{monster}) with $q = q^* \equiv 0.2445...$.  
The second result concerns the geometrical description of the 
multifractal attractor existing at $a_c$.  See ref. \cite{tsallisplastino}
for details. This gives a different method for finding a special value of $q$
which fits the results, and this value turns out to be again 0.2445.

The {\it power--law} sensitivity to the initial conditions has already been
noticed in the literature \cite{politi}.  We shall refer to it as {\it weak
sensitivity}, as opposed to the {\it exponential} law which we call {\it strong
sensitivity}.  The weak case is characteristic of {\it the edge of chaos}.  
The conclusion which seems to emerge from our work is that the various
manifestations of the edge of chaos all contain in their description a certain
parameter $q^*$, which has the same value for all of them; and that moreover
this $q^*$ is the entropic index which must be used (instead of the usual value
1) in a thermodynamic description of an edge--of--chaos system.

We shall now present the numerical work we have done, in which $q^*$ is
calculated in a third, completely different way, which involves rates of
increase of entropies.  We use the analysis developped in ref.\cite{vm} for
conservative systems.  In order to do so we partition the interval $-1 \le x
\le 1$ into $W$ equal cells; we choose (randomly or not) one of them and we
select (randomly or uniformly) $N$ values of $x$ (inside that cell) which will
be considered as initial conditions for the logistic map at a given value of
$a$.  As $t$ evolves, the $N$ points spread within the $[-1,1]$ interval in
such a way that we have a set $\{N_i(t)\}$ with $\sum_{i=1}^W N_i(t)=N\;\;
\forall t$, and a set of probabilities $\{p_i(t) \equiv N_i(t)/N\}$.
Differently from ref.\cite{vm}, we consider here the general entropy
(\ref{monster}), which for $q=1$ reduces to the entropy used in \cite{vm}.

At $t=0$, all probabilities but one are zero, hence $S_q(0)=0$. And, as $t$
evolves, $S_q(t)$ tends to increase, in all cases bounded by
$\frac{W^{1-q}-1}{1-q}$ ($\ln W$ when $q=1$), which corresponds to
equiprobability. Fluctuations are of course present 
and can be reduced by considering averages over the initial
conditions. 
As last step, we define the following rate of
increase
\begin{equation}
\kappa_q\equiv\lim_{t\rightarrow\infty}\lim_{W\rightarrow\infty}
       \lim_{N\rightarrow \infty}\frac{S_q(t)}{t}
\end{equation}
where $\kappa_1$, in the case of chaotic conservative systems, 
is expected to coincide with the standard KS 
entropy rate \cite{vm}.  Our expectations, based on the work of refs.
\cite{tsallislyra,guerberoff}, are:\\
(i) A special value $q^*$ exists such that $S_{q^*}(t)$ increases 
linearly with time. $\kappa_q$ is then {\it finite} for
$q=q^*$, {\it vanishes} for $q>q^*$ and {\it diverges} for $q<q^*$.\\
(ii) When the system is {\it strongly sensitive} to initial conditions 
($\lambda_1>0$), $q^*$ is 1 and the results of ref. \cite{vm} can be extended
to {\it dissipative} systems.\\
(iii) At the edge of chaos ($\lambda_1=0$, {\it weakly sensitive} systems), 
$q^*$ is different from $1$ and coincides with the value determined from 
the sensitivity to initial conditions (eq.(\ref{monster})) and 
from the multifractal spectrum.\\
The following results confirm these expectations.

In fig. 1 we present the case $a=2$, for which the system is chaotic (with
$\lambda_1 = \ln 2$~), and then we expect $q^*=1$. We show the time evolution
of $S_q$ for three different values of $q$~.  Only the curves for $q=1$ show a
clear linear behavior before we reach the asymptotic constant value which
characterizes the equilibrium distribution in the available part of phase
space.  The slope in the intermediate time stage does not depend on $W$ and it
is equal to the KS entropy rate $ln~2$ (for any one--dimensional system the KS
entropy is given by the positive Lyapunov exponent \cite{pesin}).  This is clear in
fig. 2 which shows $S_1(t)$ for two different cases $a=2$ and $a=1.6$~.  In
both cases the fitted slope agrees with the predicted KS entropy rates,
respectively $\ln 2$ and $0.3578$~.

So far we have shown that $q^*$ is 1 for all the cases in which the logistic
curve is chaotic, i.e. strongly sensitive to the initial conditions.  Now we
want to study the same system at its chaos threshold $a=a_c =
1.40115519...$ for which we expect $q^*=q_c = 0.2445...$.  For this value of
$a$, considerable fluctuations are observed which require an efficient and
careful averaging over the initial conditions.  Consider that the attractor
occupies only a tiny part of phase space (844 cells out of the $W=10^5$ of our
partition). We adopt the following criterion: we choose the initial
distribution (made of $N=10^6$ points) in one of the $W=10^5$ cells of the
partition and we study the number of occupied cells during the time evolution;
the integrated number of occupied cells, i.e., the sum of the numbers of
occupied cells for all time steps from iteration 1 to iteration 50, is a
measure of how good this initial condition is at spreading itself.  We repeat
the same study for each one of the $W/2$ cells in the interval $0 \le x \le 1$
and, in fig.3, we plot the integrated number of occupied cells vs. the position
of the initial distribution.  The cells for which this number is larger than a
fixed cutoff (= 5000 in fig.3) are selected for inclusion in the averaging
process (in fig.3 this is 1251 cells, out of a total $10^5$).  In fig.4 we plot
$S_q(t)$ for four different values of $q$~; the curves are an average over the
1251 initial cells selected by fig.3~.  The growth of $S_q(t)$ is found to be
linear when $q=q_c=0.2445$, while for $q<q_c$ ($q>q_c$) the curve is concave
(convex).  This behavior is similar to the one in fig.1, with a major
difference: the linear growth is not at $q=1$ (see inset(a) in fig.4), but at a
particular value of the entropic index, which happens to coincide with
$q^*=0.2445$.  In order to make this result much more convincing,
we fitted the curves $S_q(t)$ in the time
interval $[t_1,t_2]$ with the polynomial $S(t)=a+bt+ct^2$~\cite{nota2} . 
We define $
R=~|c|\cdot(t_1+t_2)/b$ as a measure of the importance of the nonlinear term in
the fit: if the points were on a perfect straight line, $R$ should be zero.  We
choose $t_1=15$ and $t_2=38$ for all $q$'s, so that the factor $(t_1+t_2)$ is
just a normalizing constant.  Fig.4(b) shows the minimum of $ R$ for
$q=q_c=0.2445$. These results are not sensitive to small changes in $t_1$ and
$t_2$. 

Precisely the same behavior, in all of its aspects, that we find here for the standard logistic map has also been found recently \cite{ananos} for its extended version where the term $x_t^2$ is generalized into $|x_t|^z$ ($z \in {\cal R}$). As expected from well known universality class considerations for this family of maps, it was found that $q^*$ depends on $z$. The same values for $q^*(z)$ were also found \cite{ananos} for a periodic family of maps which belongs to the same universality class as the logistic-like family. Finally, for the usual logistic map, the value $q^* = 0.24...$ was found through a different algorithm \cite{montangero}, closer in fact to the original definition of the Kolmogorov-Sinai entropy.
 
To summarize, we have illustrated for the logistic map the connections between
the sensitivity to initial conditions, the geometrical support in phase space,
and the linear growth in time of the entropy $S_q$.  In the case of strong
sensitivity (exponential divergence between trajectories), the geometrical
support is Euclidean, and the relevant entropy with linear growth is $S_1$, the
usual entropy.  In the case of weak sensitivity (power--law divergence), the
geometrical support is multifractal, and the relevant entropy whose growth is
linear is the non--extensive entropy $S_q$, with a special value $q^* \neq 1$
of the entropic index.  This $q^*$ is the same parameter which enters also in
the power--law divergence and in the multifractal support: the same $q^*$
describes all three phenomena.  Thus strong sensitivity and weak sensitivity to
initial conditions become unified in a single description, the difference
residing in the particular value of $q$ ( $= 1$ for the strong case). The KS
entropy rate, which is an average loss of information, is also indexed by $q$.
We believe that the scenario herein exhibited is valid for vast classes of
nonlinear dynamical systems, whose full and rigorous characterization would be
very welcome.  We conclude, as a final remark, that the still unclear
foundation of statistical mechanics on microscopic dynamics seems more than
ever to follow along the lines pioneered by Krylov \cite{krylov}. Indeed, the
crucial concept appears to be the {\it mixing} (and not only ergodicity): if
the mixing is {\it exponential} ($strong~mixing$), then $q=1$ and the standard
thermodynamical extensivity is the adequate hypothesis for those physical
phenomena, whereas at the edge of chaos the mixing is only {\it algebraic}
($weak~mixing$) and then $q \ne 1$ and thermodynamical nonextensivity is
expected to emerge.
 
One of us (C.T.) acknowledges the warm hospitality of the Dipartimento di
Fisica dell' Universit\'a di Catania where this 
work was initiated, as well as useful remarks 
from A. Politi and partial support by CNPq, 
FAPERJ and PRONEX/FINEP (Brazilian agencies).  V.L. thanks the
Blanceflor--Ludovisi Foundation and CNR for  financial support. M.B. 
acknowledges partial support from the U.S. Department of Energy (D.O.E.)
under Contract No. DE-FC02-94ER40818.


$(a)$ E-mail: latora@ct.infn.it

$(b)$ E-mail: baranger@ctp.mit.edu

$(c)$ E-mail: rapisarda@ct.infn.it

$(d)$ E-mail: tsallis@cbpf.br


\begin{figure}
\begin{center}
\epsfig{figure=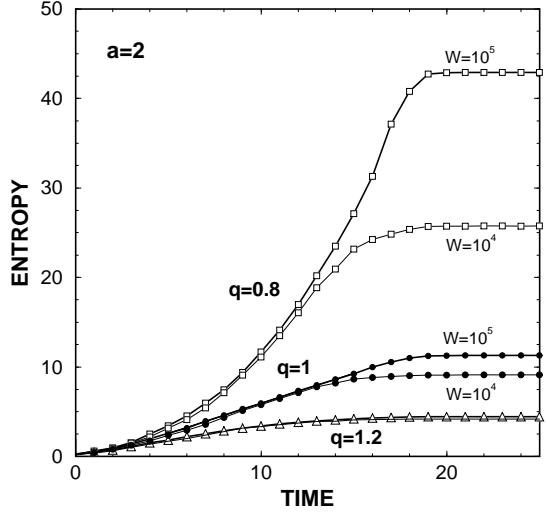,width=8truecm,angle=-90}
\end{center}
\caption{ 
Time evolution of $S_q$ for $a=2$. The interval $-1 \le x \le 1$ is
partitioned into $W$ equal cells. The initial distribution consists of
$N=10^6$ points placed at random inside a cell picked at 
random anywhere on the map.
We consider three different values of $q$ and the two cases $W=10^4$ 
and $W=10^5$.  Results are averages over 100 runs.
}
\end{figure}
\begin{figure}
\begin{center}
\epsfig{figure=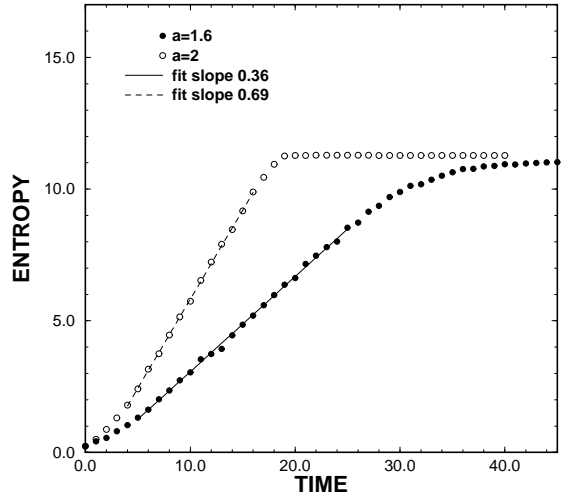,width=8truecm,angle=-90}
\end{center}
\caption{ 
Time evolution of $S_1$ for $a=2$ and $a=1.6$~; $W=10^5$ and 
averages of 100 runs. The slopes of the fits shown are respectively 
equal to the averaged Lyapunov exponents.
}
\end{figure}
\begin{figure}
\begin{center}
\epsfig{figure=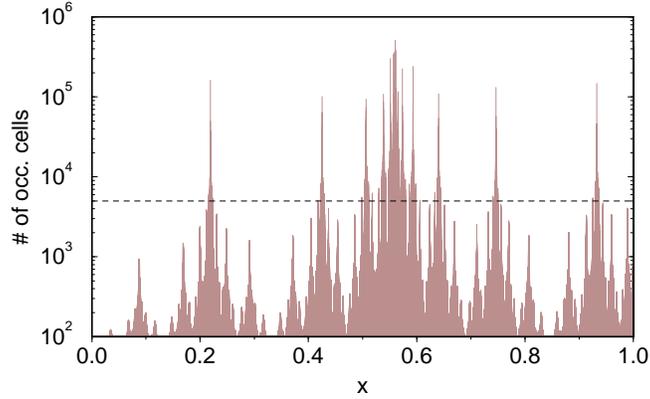,width=8truecm,angle=-90}
\end{center}
\caption{ 
Integrated number of occupied cells vs. position of the initial cell.  The
horizontal line selects the best initial conditions (see text). If we increase or decrease the value of this (numerically convenient, but dispensable) cut-off, the value for $q^*$ remains the same; what changes is the proportionality coefficient between $S_{q^*}$ and time.
}
\end{figure}
\begin{figure}
\begin{center}
\epsfig{figure=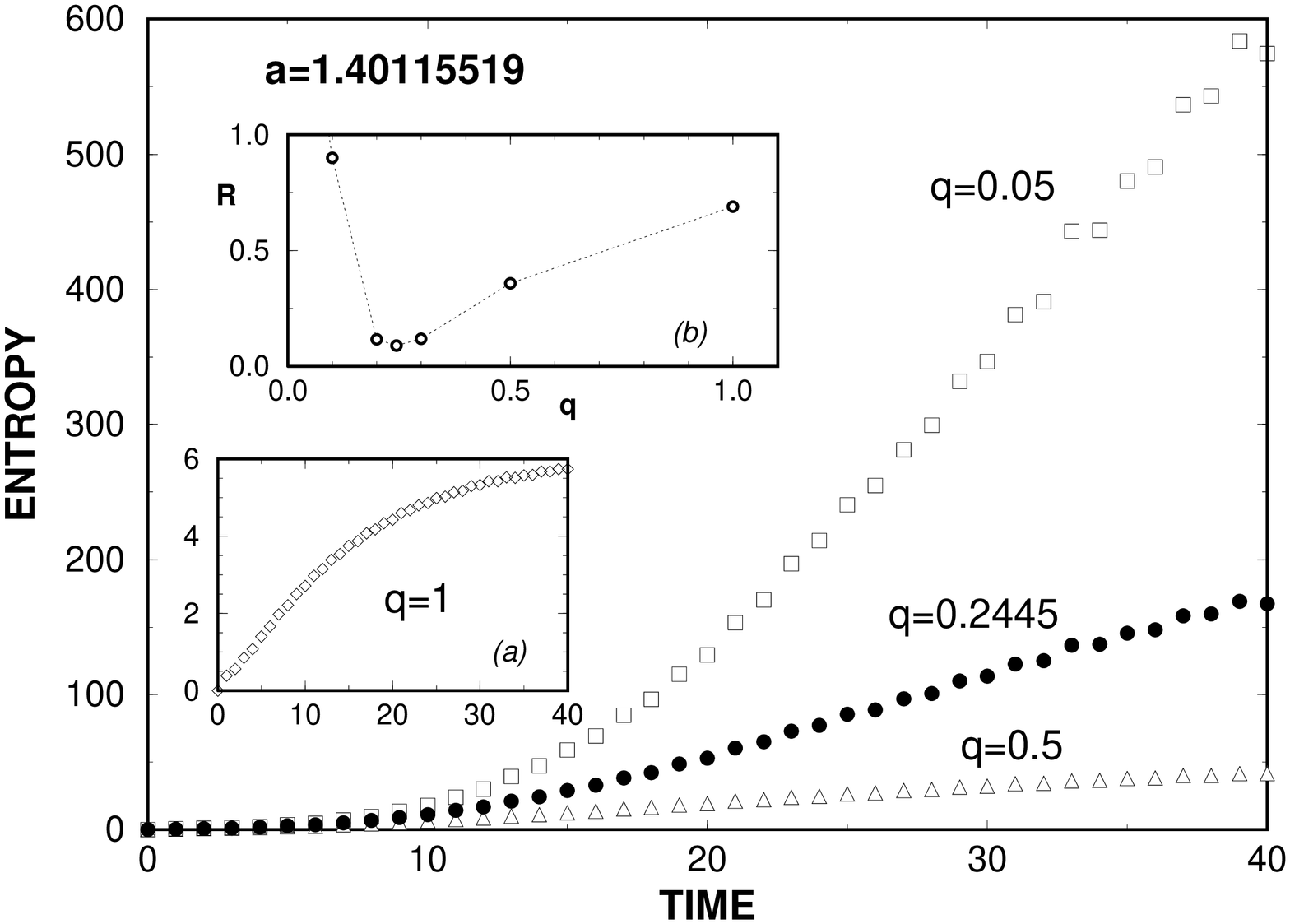,width=8truecm,angle=-90}
\end{center}
\caption{ 
Time evolution of $S_q$ for $a=a_c$~.  We consider four different values of
$q$~ and $W=10^5$. 
The case $q=1$ is reported in the inset (a) with a different scale.  
Results are averages over 1251 runs.
We show the coefficient of nonlinearity $R$ vs. $q$~ in the inset (b). 
See text. The 4-digit precision for $q^*$ was not attained through the present numerical procedure, but using the scaling $1/(1-q) = 1/\alpha_{min} - 1/\alpha_{max}$. The present procedure does not provide higher precision than $q^* = 0.24...$.
}
\end{figure}


\vfill

\begin{thebibliography}{99}
\bibitem{nota1}
At the chaos threshold the separation has a very 
fluctuating behavior.  The time series is a set of scattered points organized 
in a regular but complicated way. 
Eq. (\ref{monster}) describes the 
upper bound of this complicated structure, i.e. the quickest 
possible mixing. See details in ref.\cite{tsallislyra}.
\bibitem{nota2} 
This method is more sensitive than the standard correlation 
coefficient of the linear regression analysis in determining the best
linear slope among different values of q.
\end{thebibliography}
\end{document}